# Molecular dynamics simulations of active Brownian particles in dilute suspension: diffusion in free space and distribution in confinement

Liya Wang

*Department of Mathematics, Hong Kong University of Science and Technology, Clear Water Bay, Kowloon, Hong Kong, China*

Xinpeng Xu

*Faculty of Physics, Guangdong-Technion - Israel Institute of Technology, Shantou, Guangdong, China*

Zhigang Li

*Department of Mechanical and Aerospace Engineering, Hong Kong University of Science and Technology, Clear Water Bay, Kowloon, Hong Kong, China*

Tiezheng Qian*

*Department of Mathematics, Hong Kong University of Science and Technology, Clear Water Bay, Kowloon, Hong Kong, China*

*To whom correspondence should be addressed: maqian@ust.hk

**Abstract**

In this work, we report a new method to simulate active Brownian particles (ABPs) in molecular dynamics (MD) simulations. Immersed in a fluid, each ABP consists of a head particle and a spherical phantom region of fluid where the flagellum of a microswimmer takes effect. The orientation of the active particle is governed by a stochastic dynamics, with the orientational persistence time determined by the rotational diffusivity. To hydrodynamically drive the active particle as a pusher, a pair of active forces are exerted on the head particle and the phantom fluid region respectively. The active velocity measured along the particle orientation is proportional to the magnitude of the active force. The effective diffusion coefficient of the active particle is first measured in free space, showing semi-quantitative agreement with the analytical result predicted by a minimal model for ABPs. We then turn to the probability distribution of the active particle in confinement potential. We find that the stationary particle distribution undergoes an evolution from the Boltzmann-type to non-Boltzmann distribution as the orientational persistence time is increased relative to the relaxation time in the potential well. From the stationary distribution in confinement potential, the active part of the diffusion coefficient is measured and compared to that obtained in free space, showing a good semi-quantitative agreement while the orientational persistence time varies greatly relative to the relaxation time.

# 1 Introduction

Active particles are self-propelled, capable of converting energy from the environment or the food into directed motion.[1-4] Ubiquitous examples include bacteria,[5-8] motile cells [9-11] and artificial Janus particles.[12-14] Due to the constant energy supply and consumption, active particles are non-equilibrium by nature, making active suspensions intrinsically different from their passive counterparts.[7,15]

Active particles swimming at small length scale are governed by low Reynolds number hydrodynamics dominated by viscous damping.[16] For hydrodynamically interacting active particles, the fluid flow generated by one swimmer inevitably influences the motion of nearby swimmers. It has been generally accepted that hydrodynamic interactions play a crucial role in the dynamics of collective phenomena.[17-20] In addition, thermal Brownian noise originating from collisions with fluid particles also affects the motion of active particles significantly. In this sense, molecular dynamics (MD) simulations have the unique advantage in simulating active dynamics as both hydrodynamic interactions and thermal noise are naturally included.

The hydrodynamic flow induced by the activity of a microswimmer is usually described by using a force dipole.[3,20-22] Investigations of flagellated swimmers, e.g. *Escherichia coli* bacteria and *Chlamydomonas reinhardtii* algae, have confirmed this picture.[19,21-24] To model the motion of flagellated swimmers, flagellum is usually not explicitly described. Instead, a force exerted on the fluid is employed to incorporate the effect of a rotating flagellum. In the fluid particle dynamics method, a "phantom" spherical particle is used to model this effect.[20] In our MD simulations presented here, a spherical phantom region is introduced to an active particle, with the fluid particles in this region being subjected to the force exerted by the flagellum. The force dipole driving an active particle is formed by a pair of active forces, one exerted on its solid body and the other on the phantom region of fluid. As a result, the active particle can be modelled as a pusher or a puller depending on how the active forces are directed.

With hydrodynamic effects completely neglected, microswimmers are commonly described by a minimal model for active Brownian particles (ABPs),[2-4] which can effectively capture various fundamental features of microswimmers: overdamped dynamics, self-propelled motion, and thermal noises acting on the translational and rotational degrees of freedom.[25-28] In particular, spherical ABPs are widely used because of their simple shape.[12-14] Note that the shape of active particles has a significant effect on their dynamics,[29-33] but this is beyond the scope of the present work. In the minimal model for ABPs, one of the basic assumptions is the constant self-propulsion speed of each particle. However, this is not always the case, especially in a dense suspension of interacting active particles.[34-36] In our MD simulations, the ABP is simulated as a pusher that is hydrodynamically driven by a force dipole. Before this method is applied to the study of collective dynamics, first we focus on the individual behavior in the present work to confirm its validity. In an extremely dilute suspension, the hydrodynamic interactions are not dominant because they decay as $r^{-2}$.

Therefore, the stochastic orientational dynamics can be individually controlled. Furthermore, the activity of the active particles can be semi-quantitatively tuned at will.

The activity of an active particle can be quantified by its effective diffusion coefficient in free space, where the intriguing and unique out-of-equilibrium nature of the active particle is clearly demonstrated. Compared with a passive Brownian particle (PBP), an ABP executes a diffusive motion in a much longer time with a much larger length scale.[2,13,14] It has been derived that the active part of the diffusion coefficient is given by $D_A = v_A^2 \tau_r / 3$ in the minimal model for ABPs,[3] where $v_A$ is the active velocity and $\tau_r$ is the orientational persistence time. In our MD simulations, we will investigate to what extent this relation remains valid, given the presence of additional complexities in a more realistic description.

Compared with the diffusive motion in free space, the stochastic dynamics of active particles in confinement is more interesting because of its relevance to the physical and chemical properties of cells and biomolecules.[37,38] For active particles confined in an external potential, on the one hand, a Boltzmann-type distribution may still be realized under certain conditions, with the active particles behaving as "hot colloids" at a higher effective temperature.[14,39] In particular, the Boltzmann-type distribution can be analytically derived for run-and-tumble particles in one dimension.[40] On the other hand, accumulation of active particles at the confinement boundary is also observed when the persistence length (or run length) of active particles is comparable to or larger than the confinement length scale.[2] In our MD simulations, we will investigate how the particle distribution deviates from the Boltzmann-type and develops non-Boltzmann characteristics such as boundary accumulation.

In the present work, MD simulations are carried out to simulate the dynamics of ABPs that are realized as pushers driven by force dipoles. We investigate the diffusive motion in free space and the particle distribution in confinement. The paper is organized as follows. In section 2, we elaborate on how to realize an ABP as a pusher in MD simulations. We show that the axial velocity of the ABP exhibits a Gaussian distribution whose mean value is defined as the active velocity $v_A$ which increases with the active force $F_A$ linearly. In section 3, we investigate the diffusive motion of ABPs in free space. Our numerical results support the relation $D_A \propto v_A^2 \tau_r$, and reasons are presented for why the MD results deviate from the prediction of the minimal model. In section 4, we simulate and analyze the distribution of ABPs in an isotropic harmonic potential $U = kr^2/2$. Our results exhibit a clear evolution from the Boltzmann-type distribution to non-Boltzmann distribution as the dimensionless parameter $\mu k \tau_r$ is increased, where $\mu$ is the mobility and $k$ is the spring constant in $U$. In addition, our MD results for the stationary distribution show a semi-quantitative agreement with the prediction by the minimal model for ABPs. The paper is concluded in section 5.

## 2 Active Brownian particles in MD simulations

### 2.1 Simulation details

To investigate the ABP dynamics in a *dilute* suspension, MD simulations are carried out for three active particles placed in a cubic box, as shown in Fig. 1(a). The simulated system is composed of three active particles and a large number of fluid particles. In the present work, the ABP is realized and simulated as a pusher, which is schematically illustrated in Fig. 1(b). Each ABP consists a solid body and a fluid body. The solid body is made by a spherical particle representing the active particle's head. This spherical particle will also be called the head particle. The fluid body is made by a spherical phantom region of fluid that is centered at a position away from the head particle along a certain direction. The orientation of the active particle is represented by a unit vector $\mathbf{n}$ in the direction from the center of the phantom region to the center of the head particle. To realize the ABP as a pusher, a pair of active forces $F_A\mathbf{n}$ and $-F_A\mathbf{n}$ form a force dipole, and are applied on the head particle and the phantom region of fluid respectively. Physically, there is a thin flagellar bundle that is attached to the head particle and exerts a force on the phantom region of fluid.

The fluid particles are spherical and interact with each other through the Lennard-Jones (LJ) potential

$$V_{LJ}^{ff}(r) = 4\varepsilon_{ff}\left[\left(\frac{\sigma_{ff}}{r}\right)^{12} - \left(\frac{\sigma_{ff}}{r}\right)^{6}\right], \tag{1}$$

where $r$ is the distance between particles, and $\varepsilon_{ff}$ and $\sigma_{ff}$ denote the energy and length scales, respectively. Note that all the results in this paper are to be presented in the reduced units, with length measured by $\sigma_{ff}$, energy by $\varepsilon_{ff}$, mass by $m_f$ which is the mass of each fluid particle, and time by $\tau_0 = \sqrt{m_f \sigma_{ff}^2 / \varepsilon_{ff}}$. The LJ potential between fluid particles is cut off at $r_{cut}^{ff} = 2.5\sigma_{ff}$. In our simulations, the average number density of fluid particles is $\rho = 0.8\sigma_{ff}^{-3}$. The interaction between head particles is modelled by using the purely repulsive Weeks-Chandler-Anderson (WCA) potential,[41] which is obtained from the standard LJ potential with a truncation at the minimum potential energy at the distance $2^{1/6}\sigma_{aa}$ and an upward shift by the energy $\varepsilon_{aa}$:

$$V_{WCA}^{aa}(r) = \begin{cases} 4\varepsilon_{aa}\left[\left(\frac{\sigma_{aa}}{r}\right)^{12} - \left(\frac{\sigma_{aa}}{r}\right)^{6}\right] + \varepsilon_{aa}, & r \leq 2^{1/6}\sigma_{aa} \\ 0, & r > 2^{1/6}\sigma_{aa} \end{cases}. \tag{2}$$

A strong repulsion with the energy scale $\varepsilon_{aa} = 10\varepsilon_{ff}$ is used for head particles and the corresponding length scale is $\sigma_{aa} = 3\sigma_{ff}$. The mass of each head particle is $m_a = 10.64m_f$. The interaction between head and fluid particles is modeled by using another LJ potential with the energy scale $\varepsilon_{af} = \varepsilon_{ff}$ and length scale $\sigma_{af} = (\sigma_{aa} + \sigma_{ff})/2 = 2\sigma_{ff}$. This interaction is cut off at $r_{cut}^{af} = 2.5\sigma_{af}$.

For each active particle, the center of the fluid phantom region is away from the center of the head particle by a fixed distance of $5\sigma_{ff}$. The radius of the phantom region is $2\sigma_{ff}$ and there are always about 27 fluid particles in this region. In addition, the active force $F_A$ exerted on the phantom region is equally divided by the fluid particles in the region.

MD simulations have been carried out using the LAMMPS package.[42] The equations of motion are integrated using Velocity-Verlet algorithm with a time step of $0.0025\tau_0$. Using a Langevin thermostat in an NVE ensemble, the temperature of the fluid is controlled at $1.5\varepsilon_{ff}/k_B$, with $k_B$ being the Boltzmann constant. For the average number density of fluid particles $\rho=0.8\sigma_{ff}^{-3}$ used here, there are 108000 fluid particles placed in a simulation box measuring $L_x \times L_y \times L_z = 51.3\sigma_{ff} \times 51.3\sigma_{ff} \times 51.3\sigma_{ff}$. Periodic boundary conditions are applied in all the three directions. As a result, the spherical phantom region of an active particle may fall into several parts at the boundaries. As shown in Fig. 1(a), the phantom region of one active particle is separated into four parts at the boundary lines.

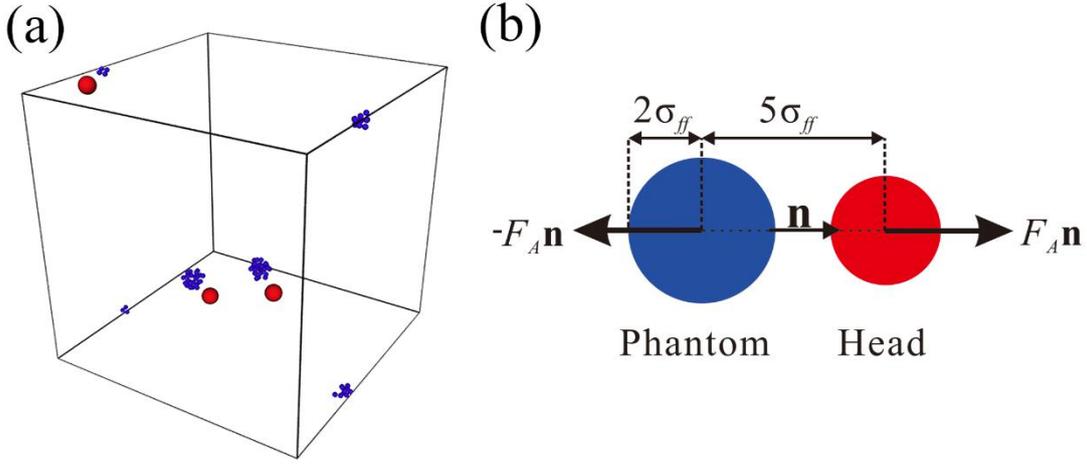

**Fig. 1** (a) A snapshot of the simulation showing a dilute suspension of three active particles in the simulation box. The red particles are the head particles and the blue particles are the fluid particles in the phantom regions. Fluid particles out of the phantom regions are not shown here. Due to the periodic boundary conditions, the fluid body of one active particle is separated into four parts. (b) The ABP modelled in this work. A force dipole is exerted on the head particle and the phantom region of fluid to model a pusher.

## 2.2 Elementary aspects of active Brownian particles

Many different models have been proposed to describe the self-propelled motion of active particles.[39,43,44] A common feature of these models is that an active particle moves under the influence of certain directional control of stochastic nature. Different from a PBP with decoupled rotational and translational motions, the self-propelled motion of an ABP result in the coupling between rotational and translational degrees of freedom.[2]

### 2.2.1  Stochastic orientational dynamics

The dynamics of a spherical ABP is governed by the overdamped Langevin equations

$$\dot{\mathbf{r}} = v_A \mathbf{n} + \sqrt{2D_T}\boldsymbol{\xi}, \tag{3}$$

and
$$\dot{\mathbf{n}} = \sqrt{2D_r}\,\mathbf{n} \times \boldsymbol{\zeta}, \tag{4}$$

in which $\mathbf{r}$ is the particle position, $\mathbf{n}$ is the unit vector denoting the particle orientation, $v_A \mathbf{n}$ is the active velocity in the direction of $\mathbf{n}$ with $v_A$ being the constant speed, $D_T$ and $D_r$ are the translational and rotational diffusion coefficients respectively, and $\boldsymbol{\xi}$ and $\boldsymbol{\zeta}$ are three-dimensional translational and rotational Gaussian white noises, with each component being of zero mean and unit variance.

In the present work, the orientational dynamics of an ABP, i.e., the time evolution of $\mathbf{n}$, is obtained by solving Eq. (4). This is accomplished as follows. Firstly, for a given value of $D_r$, Eq. (4) is numerically solved using a Python code to generate a time series of $\mathbf{n}$. Secondly, this series of $\mathbf{n}$ are used as the input to the MD simulation carried out by the LAMMPS package. The particle orientation $\mathbf{n}$ is read at each time step. Once $\mathbf{n}$ is given at a particular time step, the phantom region of fluid is located (relative to the head particle) and the force dipole along the particle orientation is then exerted, as illustrated in Fig. 1(b). As a result, the active particle is self-propelled in the direction of $\mathbf{n}$ amid the noises acting on the translational and rotational degrees of freedom.

### 2.2.2 Orientational persistence time

For an ABP whose orientation is governed by Eq. (4), it can be numerically verified that the orientational time correlation function $C(t)$ can be expressed by an exponential function

$$C(t) = \langle \mathbf{n}(t) \cdot \mathbf{n}(0) \rangle = e^{-t/\tau_r}, \tag{5}$$

where $\tau_r$ is the orientational persistence time. For spherical particles in three dimensional space, the orientational persistence time $\tau_r$ is directly related to the rotational diffusivity $D_r$ through the relation[3]

$$\tau_r = \frac{1}{2D_r}. \tag{6}$$

In Fig. 2(a), an ABP with $D_r = 0.05\tau_0^{-1}$ is taken as an example to show $C(t)$ as a function of time. It is readily seen that the numerical result for $C(t)$ can be fitted by Eq. (5) with $\tau_r = 10.32\tau_0$, which is related to $D_r$ via Eq. (6) within statistical error.

In Fig. 2(b), the product of $D_r$ and $\tau_r$ is plotted for different values of $D_r$. It is seen that the numerical value fluctuates around the theoretical value 1/2. For each value of $D_r$, $C(t)$ is calculated for five times to obtain the error bar. It is noted that for $D_r < 0.005\tau_0^{-1}$, we have $\tau_r > 100\tau_0$. Limited by the sampling time for obtaining $C(t)$, we have large statistical error for long persistence time.

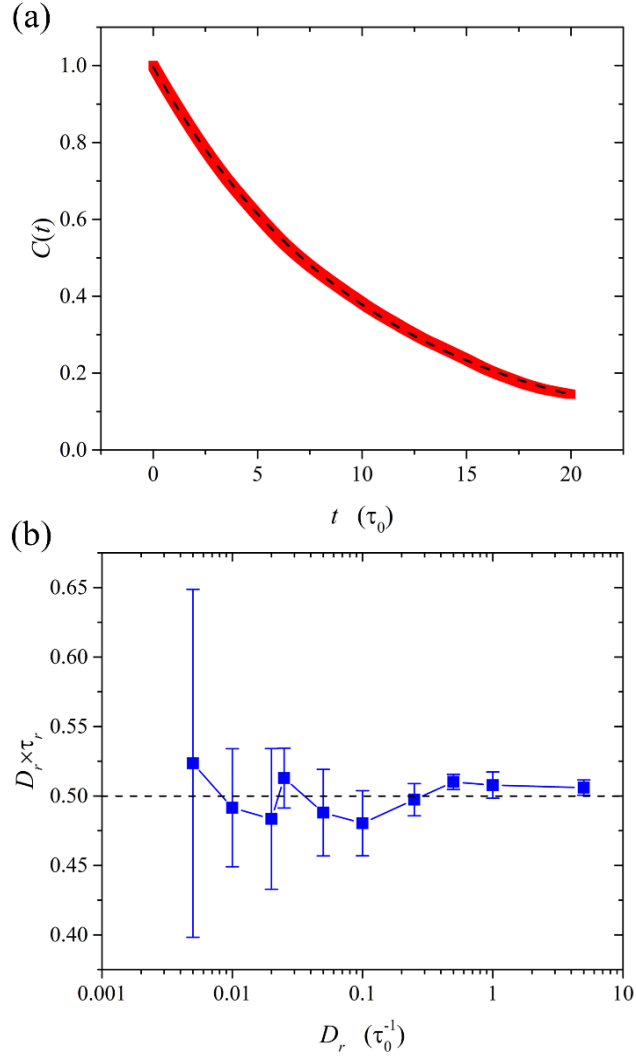

**Fig. 2** (a) Exponential decay of the orientational time correlation function $C(t)$ for $D_r = 0.05\tau_0^{-1}$. The red solid line represents the numerical result and the black dashed line represents the exponential fitting with $\tau_r = 10.32\tau_0$. (b) The product $D_r \times \tau_r$ plotted for different values of $D_r$ in the unit of $\tau_0^{-1}$. Note that for small $D_r$, $\tau_r$ is large and leads to large statistical error in a limited time duration.

Finally, we would like to point out that in our MD simulations, the particle orientation is directly taken from **n** which is obtained by solving Eq. (4). Therefore, although the active particle is surrounded by fluid particles, its orientation is not affected by the collisions with fluid particles, and the orientational time correlation function is solely controlled by the parameter $D_r$. In our MD simulations, the force dipole is exerted on the ABP in the direction of **n**, along which the active velocity is acquired and to be measured. Note that in the presence of thermal noises, the instantaneous velocity of the active particle is by no means in the direction of **n**, making a trajectory that is strongly fluctuating in time.

### 2.2.3   Active velocity

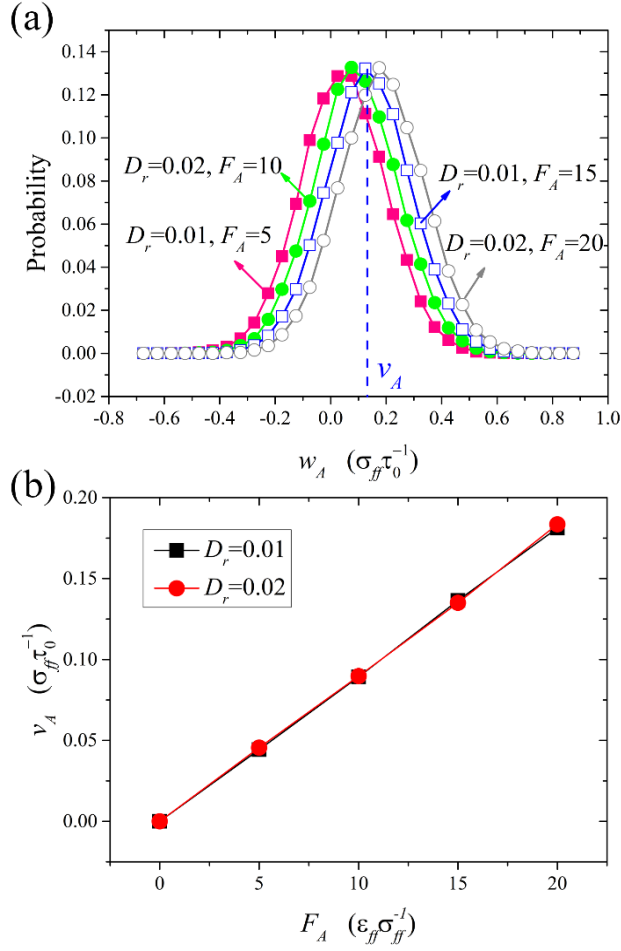

**Fig. 3** (a) Gaussian distribution of the axial velocity $w_A$, plotted for different values of rotational diffusivity $D_r$ (in the unit of $\tau_0^{-1}$) and applied force $F_A$ (in the unit of $\varepsilon_{ff}\sigma_{ff}^{-1}$). (b) The active velocity $v_A$ plotted as a function of the applied force $F_A$ for $D_r = 0.01\tau_0^{-1}$ and $0.02\tau_0^{-1}$.

Besides the orientational persistence time, the active velocity in the direction of particle orientation is another key characteristic that controls the ABP dynamics. In the commonly used minimal model described by Eqs. (3) and (4), the active velocity in the direction of $\mathbf{n}$ has a constant magnitude. However, this is no longer the case in our MD simulations. For an active particle driven by the force dipole applied in the direction of $\mathbf{n}$, the axial velocity $w_A \equiv \dot{\mathbf{r}} \cdot \mathbf{n}$ is measured at each time step in the same direction. Due to the frequent collisions with surrounding fluid particles, the axial velocity exhibits a Gaussian distribution, as shown in Fig. 3(a). Some interesting observations on the axial velocity are summarized as follows:

(i) The axial velocity $w_A$ exhibits standard Gaussian distribution. Its mean and variance are called the active velocity $v_A$ and axial velocity variance $\sigma_A^2$, respectively.

(ii) Once the size of the active particle is fixed, the active velocity $v_A$ increases with the increasing force $F_A$ linearly, as shown in Fig. 3(b). Furthermore, $v_A$ is found to be independent of $D_r$ or $\tau_r$.

(iii) The magnitude of $v_A$ measured in MD simulations is consistent with the hydrodynamic estimation. Mathematically, the flow induced at position $\mathbf{r}$ measured from the force dipole $\mathbf{p} = p\mathbf{n}$ is given by[3,21]

$$\mathbf{u}(\mathbf{r}) = \frac{p}{8\pi\eta r^3}\left[3\cos^2\theta - 1\right]\mathbf{r}, \qquad (7)$$

in which $\eta$ is the shear viscosity and $\theta$ is the angle between $\mathbf{r}$ and $\mathbf{n}$. Fig. 3(b) shows that the ratio of $v_A$ to $F_A$ is $\approx 0.01\tau_0 m_f^{-1}$, which quantitatively agrees with $v_A = F_A \times 5\sigma_{ff} \times 2/8\pi\eta l^2$ for $\theta = 0$, in which $5\sigma_{ff}$ is the distance between the center of the head particle and the center of the fluid phantom region (see Fig. 1(b)), $\eta = 2.5 m_f \sigma_{ff}^{-1}\tau_0^{-1}$ is used for the viscosity, and $l = 4\sigma_{ff}$ is used for the typical length scale. It is therefore evident that in our MD simulations, the ABP swims in the surrounding fluid as a pusher.

(iv) Once the size of the active particle is fixed, $\sigma_A^2$ is independent of $F_A$.

Based on the above observations, it can be concluded that the active velocity is instantaneously induced by the applied force dipole (given by $F_A \times 5\sigma_{ff}$ here), with its magnitude predicted by Eq. (7) semi-quantitatively. Furthermore, the active particle is subject to the random force dipole originating from the collisions with fluid particles, and hence shows a variance in the distribution of axial velocity.

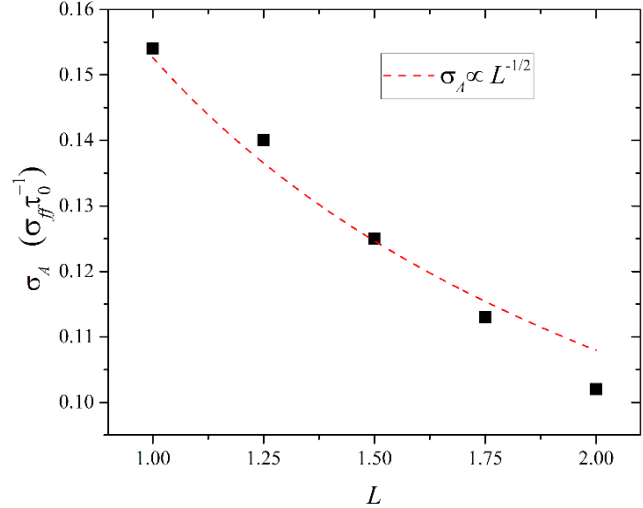

**Fig. 4** Dependence of the standard deviation $\sigma_A$ on the size of ABP.

For swimming organisms such as cells and bacteria, the size of a swimmer is typically larger than that of the fluid particle by several orders of magnitude. As a result, the ratio of the standard deviation $\sigma_A$ to the active velocity $v_A$ is negligible. In our MD simulations, however, the active particle is not that big compared to the fluid particles (see Fig. 1(b)), and hence the standard deviation $\sigma_A$ becomes appreciably large. It is interesting to observe how $\sigma_A$ would scale with the size of ABP. For this purpose, a length parameter $L$ is introduced to measure the size of ABP, with $L=1$ corresponding to the ABP illustrated in Fig. 1(b). Note that when $L$ is increased, all the length parameters are increased in proportion, including the length scale in the WCA potential for head particles, the radius of the fluid phantom region, and the distance between the head particle and the fluid phantom region. Fig. 4 shows that the standard deviation $\sigma_A$ does decrease with the increasing $L$. A theoretical argument can be made to predict $\sigma_A \propto L^{-1/2}$, which is indeed indicated by Fig. 4.

### 2.2.4 Critical persistence length

In the present work, the particle orientation **n** evolves according to Eq. (4) and is supplied to the MD simulations as external input. Theoretically, any orientational persistence time $\tau_r$ can be used. However, if $\tau_r$ is very short, then the effect of the active velocity $v_A \mathbf{n}$ in Eq. (3) becomes that of the white noises with no time correlation. In our MD simulations, we find that to make the active particles behave truly 'actively', the persistence time $\tau_r$ has to be sufficiently large. While we understand that this is only from a phenomenological perspective, we believe that a sufficiently large $\tau_r$ is necessary for the ABPs to behave differently from PBPs. This will be made evident in the next two sections.

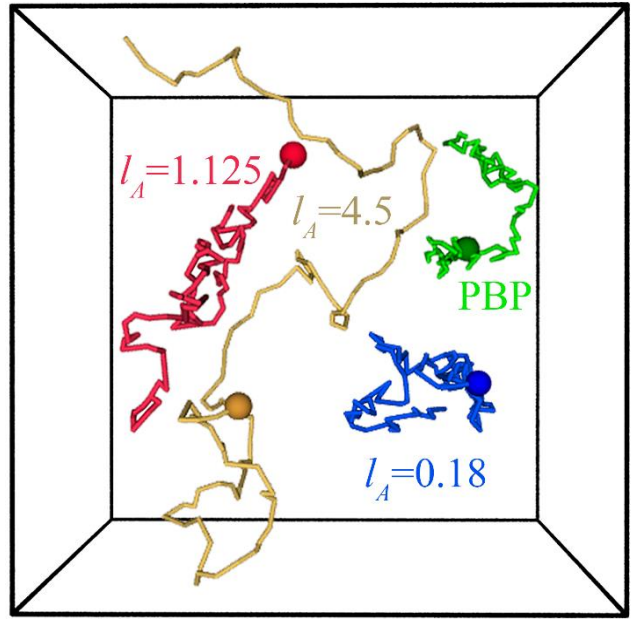

**Fig. 5** Trajectories of a PBP and three ABPs with $l_A = 0.18\sigma_{ff}$ from $\tau_r = 1\tau_0$, $v_A = 0.18\sigma_{ff}\tau_0^{-1}$, $l_A = 1.125\sigma_{ff}$ from $\tau_r = 25\tau_0$, $v_A = 0.045\sigma_{ff}\tau_0^{-1}$, and $l_A = 4.5\sigma_{ff}$ from $\tau_r = 50\tau_0$, $v_A = 0.09\sigma_{ff}\tau_0^{-1}$. Each trajectory includes 100 frames with a time duration of $25\tau_0$.

Fig. 5 shows the trajectories for different values of the persistence length $l_A$ defined by

$$l_A = v_A \tau_r, \qquad (8)$$

which quantifies the step length for an active particle's random walk. It is seen that the trajectory of the ABP of $l_A = 0.18\sigma_{ff}$ exhibits a random walk of very short step length, which is very much close to that of a PBP. In this case the persistence time $\tau_r = 1\tau_0$ is very short. Although a large active velocity $v_A = 0.18\sigma_{ff}\tau_0^{-1}$ is used, the persistence length $l_A = 0.18\sigma_{ff}$ is still very short and the active particle shows no appreciable difference from a PBP. As the persistence length is increased to $l_A = 4.5\sigma_{ff}$, the trajectory is formed by a sequence of 'straight' lines. In this case the ABP behaves truly actively, capable of exploring a much larger space compared to the PBP.

Through our simulations, we find that the self-propelled motion of ABP can be effectively distinguished from the random walk of PBP if the persistence length exceeds $l_A = 1.125\sigma_{ff}$, as illustrated in Fig. 5. The critical persistence length is therefore taken at $\approx 1\sigma_{ff}$. In the results presented below, $l_A = 1.125\sigma_{ff}$ serves as the lower bound for all the active particles.

## 3  Active Brownian particles in free space

Both PBPs and ABPs exhibit ballistic motion at short time scales, but enter into the regime of diffusive motion at long time scales.[13] Here we measure the effective diffusivity for ABPs in free space without confining potential.

We start from the diffusivity of a PBP for reference. The PBP used for this purpose is just the head particle. No phantom region of fluid is introduced and no external force dipole is applied either. To reduce the statistical fluctuations, simulations have been performed for many times with different initial conditions. From the mean square displacement (MSD) which increases with time linearly as shown in Fig. 6(a), we obtain the translational diffusion coefficient $D_T$ from

$$D_T = \frac{\left\langle \left| \mathbf{r}(t) - \mathbf{r}(0) \right|^2 \right\rangle_{PBP}}{6t}. \tag{9}$$

Through a linear fitting, $D_T$ is found to be $0.015\sigma_{ff}^2\tau_0^{-1}$. According to the Einstein relation $\gamma D_T = k_B T$, the drag coefficient $\gamma$ is approximately $100 m_f \tau_0^{-1}$. Using the Stokes drag coefficient $\gamma = 6\pi\eta R$ for a spherical particle and $\eta = 2.5 m_f \sigma_{ff}^{-1} \tau_0^{-1}$ for the viscosity, it is estimated that the radius of the PBP is about $2\sigma_{ff}$. This is consistent with the fact that the interaction between head and fluid particles is modeled by a LJ potential with length scale $\sigma_{af} = (\sigma_{aa} + \sigma_{ff})/2 = 2\sigma_{ff}$.

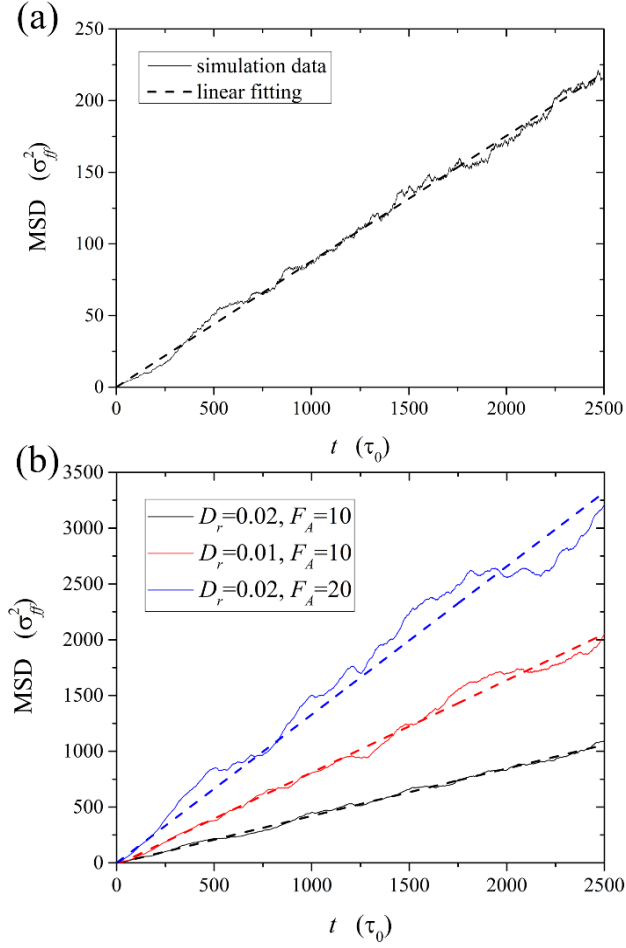

**Fig. 6** (a) MSD for PBPs (solid line), with $D_T$ found to be $0.015\sigma_{ff}^2\tau_0^{-1}$ through a linear fitting (dashed line). (b) MSD for three ABPs with different values of $D_r$ (in the unit of $\tau_0^{-1}$) and $F_A$ (in the unit of $\varepsilon_{ff}\sigma_{ff}^{-1}$). In each case, the solid line represents the simulation data and the dashed line of the same color represents the corresponding linear fitting.

Now we turn to ABPs with $D_r = 0.01\tau_0^{-1}$ and $0.02\tau_0^{-1}$ subject to several different values of the active force $F_A$. To reduce the statistical fluctuations, simulations have been performed for many times with different initial conditions and different orientational trajectories of $\mathbf{n}$. Fig. 6(b) shows the MSD results for three representative ABPs in a time duration of $2500\tau_0$. Although the MSD lines do exhibit a linear increase with time, it is noted that statistical fluctuations are amplified by the increase of the effective diffusivity itself. This is due to the insufficiency of sampling in the limited time duration. Compared with the PBP result in Fig. 6(a), Fig. 6(b) shows that the diffusive motion of ABPs is much faster, with a much larger effective diffusion coefficient $D_E$ defined by

$$D_E = \frac{\langle |\mathbf{r}(t) - \mathbf{r}(0)|^2 \rangle_{ABP}}{6t}. \qquad (10)$$

Subtracting the passive part $D_T$ from $D_E$, we obtain the active diffusion coefficient $D_A \equiv D_E - D_T$, which measures the contribution of self-propelled motion to diffusion. For the ABPs modelled by Eqs. (3) and (4), $D_A$ is given by[3]

$$D_A = \frac{v_A^2 \tau_r}{3}. \tag{11}$$

While Eqs. (3) and (4) only describe the ABPs in a minimal model, it is still interesting to see if the MD results obtained here support $D_A \propto v_A^2 \tau_r$.

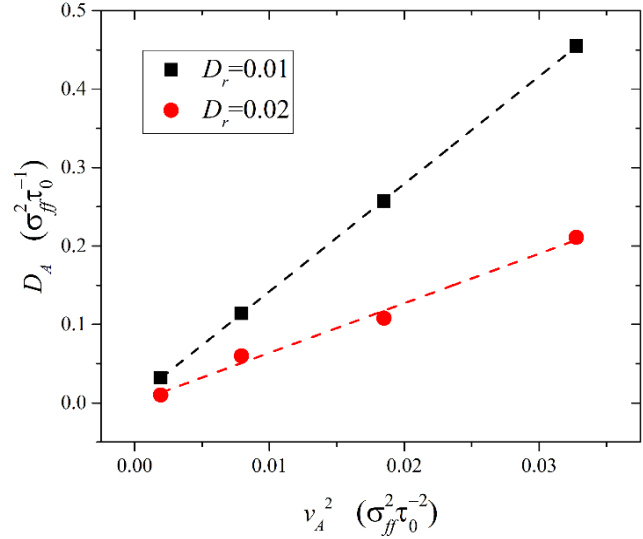

**Fig. 7** MD simulation results for the dependence of $D_A$ on $v_A^2$ for $D_r = 0.01 \tau_0^{-1}$ and $0.02 \tau_0^{-1}$. From the linear fitting (dashed lines), the prefactor $\alpha$ in $D_A \simeq \alpha v_A^2 \tau_r$ is found to be 0.274 for $D_r = 0.01 \tau_0^{-1}$ and 0.252 for $D_r = 0.02 \tau_0^{-1}$.

Fig. 7 shows the dependence of $D_A$ on $v_A^2$ for two different values of $D_r$. It is seen that for a given $D_r$, $D_A$ increases with $v_A^2$ linearly. Furthermore, the slope for $D_r = 0.01 \tau_0^{-1}$ is approximately twice as big as that for $D_r = 0.02 \tau_0^{-1}$, indicating $D_A \propto v_A^2 \tau_r$ with $\tau_r = 1/2D_r$. Although our MD simulation results support $D_A \propto v_A^2 \tau_r$, the prefactor $\alpha$ in $D_A \simeq \alpha v_A^2 \tau_r$ is found to be less than $1/3$ predicted by the minimal model in Eq. (11). This may be attributed to the additional complexities which are inherent in our MD simulations and beyond the description by Eqs. (3) and (4). These include:

(i) Due to the frequent collisions of the active particle with surrounding fluid particles, the axial velocity $w_A$ exhibits a Gaussian distribution, with the active velocity $v_A$ defined as the mean of $w_A$ (see Fig. 3). The use of a constant $v_A$ in Eq. (3) is an oversimplification.

(ii) The orientational persistence time $\tau_r$ is associated with the time evolution of particle orientation $\mathbf{n}$. Although the force dipole is applied in the direction of $\mathbf{n}$, the induced velocity may deviate from this direction due to various noises in MD simulations. The use of $v_A \mathbf{n}$ in Eq. (3) is an oversimplification again.

(iii) According to the way the ABP is constructed (see Fig. 1(b)), the active particle is by no means spherical and the use of $D_T$ for a spherical particle in Eq. (3) is an oversimplification yet again.

# 4 Active Brownian particles in confinement

In this section we investigate the distribution of non-interacting ABPs confined by an isotropic harmonic potential $U(r) = kr^2/2$, with the distance $r$ measured from the center of the simulation box. A series of different values for the spring constant $k$ (in the unit of $\varepsilon_{ff}\sigma_{ff}^{-2}$) will be used to explore different regimes of confinement. Time averaging is performed over an ensemble of particle trajectories. For weaker confinement, longer time averaging is needed to remove statistical fluctuations.

## 4.1 Boltzmann distribution of passive Brownian particles

We start from the Boltzmann distribution of a PBP. The PBP used for this purpose is still the head particle, with no phantom region and no external force dipole. The equilibrium probability density function (PDF) $g(r)$ is given by the Boltzmann distribution:

$$g(r) \propto e^{-\frac{kr^2}{2k_B T}}. \tag{12}$$

Fig. 8 shows $g(r)$ as a function of $r$ for the spring constant $k = \varepsilon_{ff}\sigma_{ff}^{-2}$. Note that the PDF $g(r)$ is defined for $r \geq 0$ with the normalization condition

$$\int_0^\infty g(r) 4\pi r^2 dr = 1. \tag{13}$$

To better present $g(r)$ visually, $r$ is extended to cover $(-\infty, +\infty)$ and the data for $r \geq 0$ are mirrored to the negative half of $r$ axis, with the curve of $g(r)$ being symmetric about $r = 0$. This applies to all the $g(r)$ curves in this section. Fig. 8 shows that the numerical result for $g(r)$ is quantitatively described by Eq. (12). Such agreement is also achieved for other values of the spring constant.

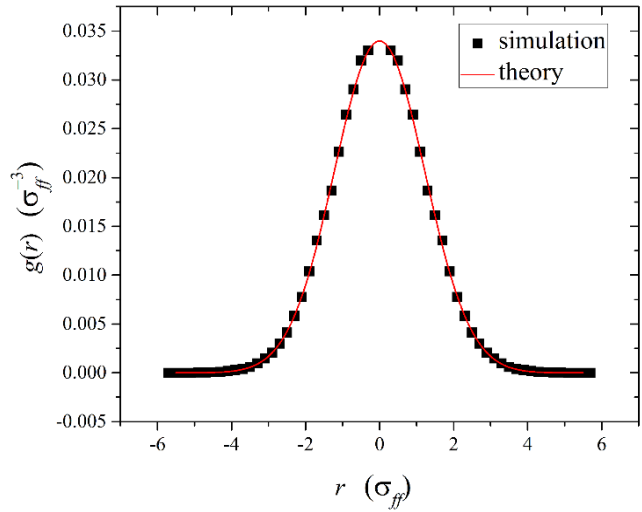

**Fig. 8** The equilibrium PDF $g(r)$ of the confined PBP for $k = \varepsilon_{ff}\sigma_{ff}^{-2}$.

## 4.2 Stationary distribution of active Brownian particles

Now we consider an ABP confined by the harmonic potential $U(r) = kr^2/2$ and focus on its stationary PDF $g(r)$. To understand the physical picture, we start from a minimal model described by Eq. (4) and

$$\dot{\mathbf{r}} = -\mu \nabla U + v_A \mathbf{n} + \sqrt{2D_T}\,\xi, \tag{14}$$

in which $\mu$ is the mobility coefficient, which is the inverse of the drag coefficient $\gamma$. Compared with Eq. (3), Eq. (14) includes an additional term $-\mu \nabla U$ due to the confining potential $U$. In the limit of $D_T \to 0$, the particle motion is confined by the boundary $r = r_B$ with $r_B$ given by

$$r_B = \frac{v_A}{\mu k}. \tag{15}$$

At finite $D_T$, the particle can still go beyond $r = r_B$ with the assistance of thermal noises. The relaxation time for the overdamped motion in the confining potential $U$ is $1/\mu k$, and the other time scale is the orientational persistence time $\tau_r$. Here we introduce a dimensionless parameter $R_1$ to measure the ratio of $\tau_r$ to $1/\mu k$:

$$R_1 = \frac{\tau_r}{1/\mu k} = \mu k \tau_r. \tag{16}$$

When $R_1 \ll 1$, the orientational persistence time $\tau_r$ is very short compared to $1/\mu k$, and the active term $v_A \mathbf{n}$ plays the role of white noises effectively. This leads to a Boltzmann-type distribution of ABP given by $g_B(r) \propto \exp(-\mu k r^2/2D_E)$, with the effective temperature given by $k_B T_E = D_E/\mu$ where $D_E$ is the effective diffusivity. The width of this distribution is about $\sqrt{D_E/\mu k}$. Another dimensionless parameter $R_2$ can be introduced to measure the ratio of $\sqrt{D_E/\mu k}$ to $r_B$. Using $D_E = D_T + D_A \approx D_A$ and $D_A \approx \alpha v_A^2 \tau_r$, we have

$$R_2 \approx \sqrt{\frac{v_A^2 \tau_r}{\mu k}} \Big/ \frac{v_A}{\mu k} = \sqrt{\mu k \tau_r}, \tag{17}$$

which is $\sqrt{R_1}$. It follows that when $R_1 \ll 1$, $R_2 \ll 1$ as well. This means that the Boltzmann-type distribution $g_B(r)$ can be realized and accommodated within the boundary $r = r_B$.

The fact that $R_1 \ll 1$ leads to $R_2 \ll 1$ can be regarded as a self-consistency check. From $R_1 \ll 1$, the particle activity can be effectively taken as white noises with $\tau_r \to 0$. As a result, the Boltzmann-type distribution is expected to occur at the effective temperature $T_E$. This distribution is then found to be much narrower that the boundary set by the confinement potential and the particle activity, and hence it is realizable.

The self-consistency check above also indicates that the Boltzmann-type distribution $g_B(r)$ will be invalidated by the increase of $R_1$. For $R_1 \gtrsim 1$, this distribution would meet the confinement boundary $r = r_B$ and hence can no longer be realized. In fact, when $\tau_r$ becomes comparable to $1/\mu k$, the PDF $g(r)$ shows a plateau rather than a peak around $r = 0$. When $\tau_r$ becomes much larger than $1/\mu k$, the ABP spends a short time ($\Box 1/\mu k$) travelling in the potential field but a long time ($\Box \tau_r$) staying near the boundary $r = r_B$ before turning around. This corresponds to the accumulation of probability at the confinement boundary, with the PDF $g(r)$ showing a bimodal distribution peaked near $r = \pm r_B$. The larger $R_1$ is, the sharper the peak is.

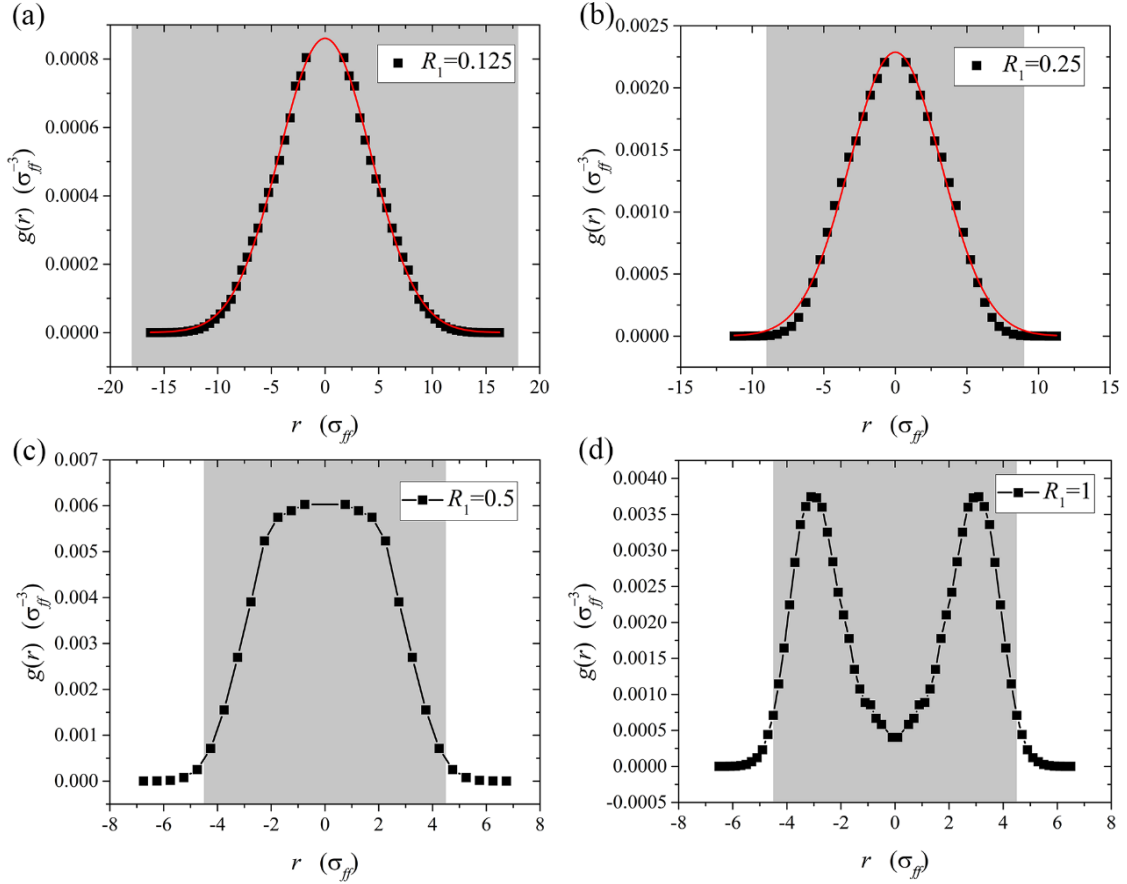

**Fig. 9** Evolution of the PDF $g(r)$ with the increase of $R_1$. (a) The Boltzmann-type distribution for $R_1 = 0.125$. (b) A distribution slightly deviating from the Boltzmann-type for $R_1 = 0.25$. (c) A distribution exhibiting a plateau in the central region for $R_1 = 0.5$. (d) A bimodal distribution for $R_1 = 1$ with the accumulation of probability near $r = \pm r_B$. Here the red line represents a fitting of the Boltzmann-type and the gray region is bounded by $r = \pm r_B$.

In consistency with the above discussion, our MD simulations have shown how the stationary PDF $g(r)$ evolves with the change of $R_1$. The same $\tau_r = 25\tau_0$ is used for all the four cases presented below. Fig. 9(a), (b) and (c) are produced for an ABP with $v_A = 0.09\sigma_{ff}\tau_0^{-1}$ using different values of $k$. Limited by the computational capability, it is unrealistic to have a confinement potential with a very small value of $k$ and hence a very wide distribution. Fig. 9(a) shows the case for $k = 0.5\varepsilon_{ff}\sigma_{ff}^{-2}$, from which we have $R_1 = 0.125$ and $r_B = 18\sigma_{ff}$ using the drag coefficient $\gamma = 100 m_f \tau_0^{-1}$ and $\mu = 1/\gamma$. This is the smallest value of $R_1$ we can access. The Boltzmann-type distribution is well maintained at this value of $R_1$. Here the red line represents a fitting of the Boltzmann-type and the gray region is bounded by $r = \pm r_B$. In this case, the particle distribution is completely within the confinement boundary. Fig. 9(b) shows the case for $k = \varepsilon_{ff}\sigma_{ff}^{-2}$, $R_1 = 0.25$, and $r_B = 9\sigma_{ff}$. In this case, the particle distribution touches the confinement boundary and hence slightly deviate from the Boltzmann-type distribution. Fig. 9(c) shows the case for $k = 2\varepsilon_{ff}\sigma_{ff}^{-2}$, $R_1 = 0.5$, and $r_B = 4.5\sigma_{ff}$. In this case, the particle distribution is appreciably different from the Boltzmann-type and exhibits a plateau in the central region. Finally, Fig. 9(d) shows the case for $k = 4\varepsilon_{ff}\sigma_{ff}^{-2}$, $R_1 = 1$, and $r_B = 4.5\sigma_{ff}$. Here a larger active velocity $v_A = 0.18\sigma_{ff}\tau_0^{-1}$ is used to ensure that $r_B$ is not too small. This is to avoid the domination of thermal noises in a very narrow region. In this case, the particle distribution is bimodal, corresponding to the accumulation of probability near $r = \pm r_B$.

## 4.3 A quantitative analysis for the stationary distribution

A quantitative analysis can be carried out by focusing on the stochastic dynamics of the $x$ coordinate. The $x$ component of Eq. (14) can be written as

$$\dot{x}(t) = -\mu k x(t) + \sqrt{2D_A}\varphi(t) + \sqrt{2D_T}\xi_x, \tag{18}$$

where $\varphi$ is a colored noise due to the active motion and $\xi_x$ is the zero-mean unit-variance Gaussian white noise in the $x$ direction, with

$$\langle \varphi(t_2)\varphi(t_1) \rangle = \frac{1}{2\tau_r}e^{-|t_2-t_1|/\tau_r}, \tag{19}$$

$$\langle \xi_x(t_2)\xi_x(t_1) \rangle = \delta(t_2 - t_1). \tag{20}$$

At sufficiently large $t$, the mean square displacement of $x$ with respect to $r = 0$ can be analytically expressed as

$$\langle x^2 \rangle = \frac{D_A}{\mu k(\mu k\tau_r + 1)} + \frac{D_T}{\mu k}, \tag{21}$$

in which the contribution of the colored noise shows a dependence on $R_1 = \mu k\tau_r$. In the limit of $R_1 \to 0$, $\varphi$ is effectively a white noise and we have $\langle x^2 \rangle = (D_A + D_T)/\mu k = D_E/\mu k$, which is in agreement with the Boltzmann-type distribution $g_B(r) \propto \exp(-\mu k r^2/2D_E)$ discussed above. However, when $\tau_r$ becomes comparable to $1/\mu k$, the correction by $1/(\mu k\tau_r + 1)$ needs to be taken into consideration.

In our MD simulations, $\langle x^2 \rangle$ is computed according to its definition

$$\langle x^2 \rangle = \int_0^\infty x^2 f(x) dx, \tag{22}$$

where $f(x)$ is the stationary marginal PDF of the $x$ coordinate, which can be measured directly. It can also be obtained by measuring the stationary PDF $g(r)$ and performing the integration as

$$f(x) = \int_{-\infty}^{\infty} \int_{-\infty}^{\infty} g\left(\sqrt{x^2 + y^2 + z^2}\right) dy dz = \int_{0}^{\infty} g\left(\sqrt{x^2 + \rho^2}\right) 2\pi \rho d\rho. \tag{23}$$

Fig. 10 shows the marginal PDF for $R_1 = 0.5$ and $2.5$. The two sets of data are in good agreement, one directly measured in simulations, and the other obtained from $g(r)$ by the use of Eq. (23).

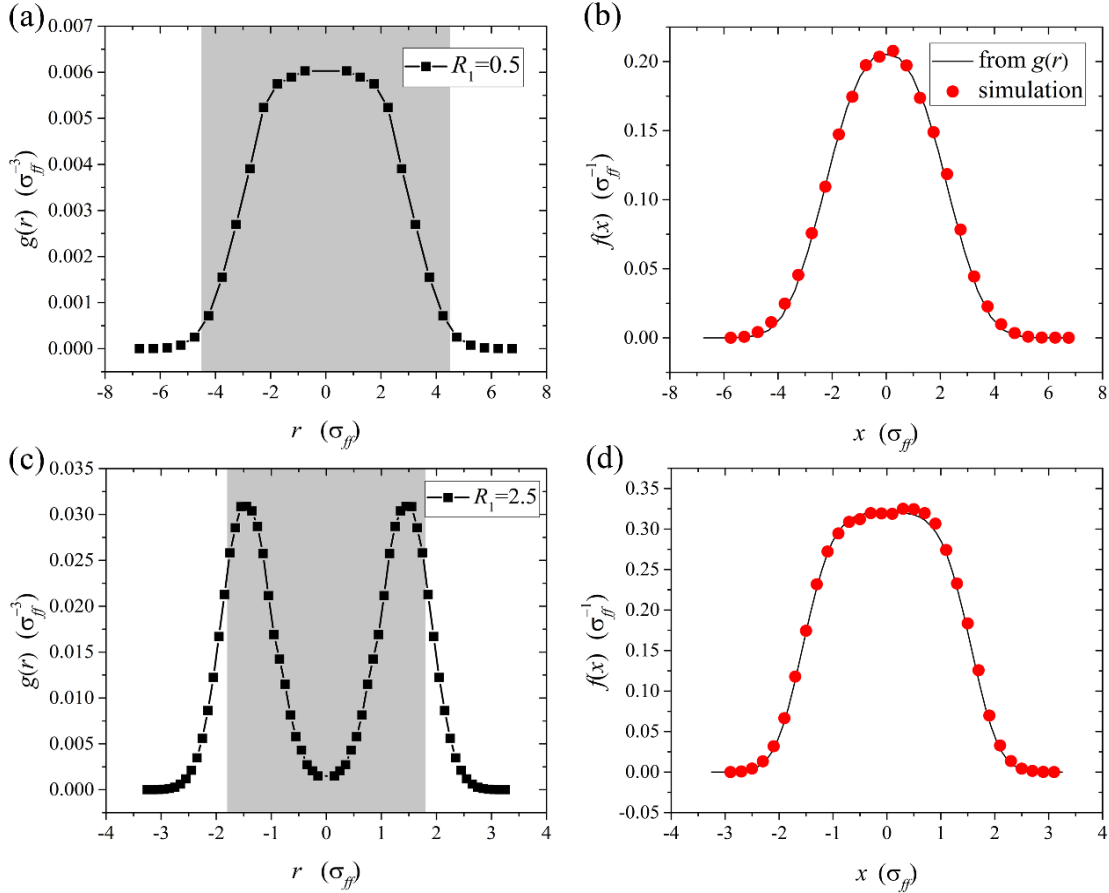

**Fig. 10** The PDF $g(r)$ and marginal PDF $f(x)$. (a) $g(r)$ for $R_1 = 0.5$. (b) $f(x)$ for $R_1 = 0.5$. (c) $g(r)$ for $R_1 = 2.5$. (d) $f(x)$ for $R_1 = 2.5$. In (b) and (d), $f(x)$ directly measured in simulations (represented by solid circles) is compared to that obtained from $g(r)$ by the use of Eq. (23) (represented by solid line), with good agreement.

We can easily compute $\langle x^2 \rangle$ by measuring $f(x)$ and using Eq. (22). We then substitute the value of $\langle x^2 \rangle$ into Eq. (21) to deduce the value of $D_A$. This involves the use of $D_T = 0.015 \sigma_{ff}^2 \tau_0^{-1}$ measured in section 3, $\mu = 1/\gamma$ with $\gamma = 100 m_f \tau_0^{-1}$ such that $\gamma D_T = k_B T = 1.5 \varepsilon_{ff}$, and $\tau_r = 1/2D_r = 25\tau_0$ with $D_r = 0.02 \tau_0^{-1}$ used in Eq. (4) for the orientational dynamics. The value of $D_A$ so deduced from $\langle x^2 \rangle$ in the confinement potential is denoted by $D_{AC}$ here. On the other hand, the active part of the diffusion coefficient $D_A \equiv D_E - D_T$, obtained by measuring $D_E$ in free space (presented in section 3), is denoted by $D_{AF}$ here. A comparison between $D_{AC}$ and $D_{AF}$ is made in Table 1 for six different values of $R_1$ from $0.125 \square 1$ to $10 \square 1$. It is readily seen that $D_{AF}$ is always less than $D_{AC}$ by about 15% to 29%. It is also interesting to note that the values of $D_{AC}$ show a better agreement with the relation $D_A \propto v_A^2$, with $0.29/0.07 \approx (0.18/0.09)^2$.

Table 1 Comparison between $D_{AC}$ and $D_{AF}$

| $R_1$ | $v_A\,(\sigma_{ff}\tau_0^{-1})$ | $\langle x^2 \rangle\,(\sigma_{ff}^2)$ | $D_{AC}\,(\sigma_{ff}^2/\tau_0)$ | $D_{AF}\,(\sigma_{ff}^2/\tau_0)$ | Error |
|---|---|---|---|---|---|
| 0.125 | 0.09 | 15.77 | 0.072 | 0.058 | 19.4% |
| 0.25 | 0.09 | 7.29 | 0.072 | 0.058 | 19.4% |
| 0.5 | 0.09 | 3.05 | 0.069 | 0.058 | 15.9% |
| 1 | 0.18 | 3.98 | 0.288 | 0.208 | 27.7% |
| 2.5 | 0.18 | 0.98 | 0.291 | 0.208 | 28.5% |
| 10 | 0.18 | 0.10 | 0.275 | 0.208 | 24.4% |

Finally, to comment on this comparison and the relation $D_A \propto v_A^2$ being better satisfied by $D_{AC}$, we would like to point out the following:

(i) As discussed at the end of section 3, the ABPs in our MD simulations involve additional complexities that are beyond the description by Eqs. (3), (4), and (14), from which $D_{AC} = D_{AF}$ is expected.

(ii) The relative difference between $D_{AC}$ and $D_{AF}$ is always smaller than 29%. This means that the minimal model based on Eqs. (3), (4), and (14) is semi-quantitatively accurate to describe the ABPs, for $R_1$ varying over two orders of magnitude (from 0.125 to 10).

(iii) In the confinement potential, the particle trajectories show better statistical convergence than in free space. This could be the reason for $D_A \propto v_A^2$ being better satisfied by $D_{AC}$.

## 5  Conclusions

In this work, we have carried out MD simulations in which ABPs are realized as pushers, each driven by a force dipole hydrodynamically. Each active particle consists of the head particle and the spherical phantom region of fluid where the flagellum takes effect. The orientation of the active particle is represented by the unit vector $\mathbf{n}$ in the direction from the center of the phantom region to the center of the head particle. To drive the active particle as a pusher, the active forces $F_A\mathbf{n}$ and $-F_A\mathbf{n}$ are applied on the head particle and the phantom region of fluid respectively. The stochastic dynamics of $\mathbf{n}$ is controlled by the rotational Gaussian white noises, and the orientational persistence time $\tau_r$ is solely determined by the rotational diffusivity $D_r$ through the relation $\tau_r = 1/2D_r$. Due to the frequent collisions of the active particle with surrounding fluid particles, the axial velocity $w_A$ of the active particle, measured in the direction of $\mathbf{n}$, exhibits a Gaussian distribution. The mean value of $w_A$ is defined as the active velocity $v_A$, which increases with the active force $F_A$ linearly.

In comparison with the overdamped ABPs described by the minimal model with constant active velocity, our MD simulations show the following results. (i) The active part of the diffusion coefficient $D_A$ measured in free space supports the relation $D_A \propto v_A^2 \tau_r$ although the proportionality constant deviates from $1/3$ predicted by the minimal model. (ii) In the isotropic harmonic potential $U = kr^2/2$, the stationary particle distribution undergoes an evolution from the Boltzmann-type distribution to non-Boltzmann distribution as the dimensionless parameter $\mu k \tau_r$ is increased. (iii) From the stationary particle distribution in the confinement potential, the active part of the diffusion coefficient can be measured and then compared to that measured in free space, with the relative difference always less than 29%. This semi-quantitative agreement is fairly good because the comparison has been carried out for $\mu k \tau_r$ varying over two orders of magnitude.

These results demonstrate that the pushers realized in our MD simulations are able to capture the salient features of the overdamped ABPs described by the minimal model. The common and convenient use of the minimal model is therefore justified on the one hand. On the other hand, the pushers in our MD simulations can interact with each other via hydrodynamic coupling, from which many interesting collective phenomena may emerge. This represents a direction to be pursued.

## Conflicts of interest

There are no conflicts of interest to declare.

## Acknowledgements

This work is supported by Hong Kong RGC CRF grant No. C1018-17G and GRF grant No. 16228216.

## References

1. S. Ramaswamy, *Annu. Rev. Condens. Matter Phys.*, 2010, **1**, 323-345.
2. C. Bechinger, R. Di Leonardo, H. Löwen, C. Reichhardt, G. Volpe and G. Volpe, *Rev. Mod. Phys.*, 2016, **88**, 045006.
3. A. Zöttl and H. Stark, *J. Phys.: Condens. Matter*, 2016, **28**, 253001.
4. P. Romanczuk, M. Bär, W. Ebeling, B. Lindner and L. Schimansky-Geier, *Eur. Phys. J. Special Topics*, 2012, **202**, 1-162.
5. H. C. Berg and R. A. Anderson, *Nature*, 1973, **245**, 380.
6. L. H. Cisneros, J. O. Kessler, S. Ganguly and R. E. Goldstein, *Phys. Rev. E*, 2011, **83**, 061907.
7. M. E. Cates, *Rep. Prog. Phys.*, 2012, **75**, 042601.
8. R. M. Harshey, *Annu. Rev. Microbiol.*, 2003, **57**, 249-273.
9. D. Selmeczi, L. Li, L. I. Pedersen, S. Nrrelykke, P. H. Hagedorn, S. Mosler, N. B. Larsen, E. C. Cox and H. Flyvbjerg, *Eur. Phys. J. Special Topics*, 2008, **157**, 1-15.
10. H. U. Boedeker, C. Beta, T. D. Frank and E. Bodenschatz, *EPL*, 2010, **90**, 28005.

11. B. M. Friedrich and F. Jülicher, *Proc. Natl. Acad. Sci. U. S. A.*, 2007, **104**, 13256-13261.
12. I. Buttinoni, J. Bialké, F. Kümmel, H. Löwen, C. Bechinger and T. Speck, *Phys. Rev. Lett.*, 2013, **110**, 238301.
13. J. Howse, R. Jones, A. Ryan, T. Gough, R. Vafabakhsh and R. Golestanian, *Phys. Rev. Lett.*, 2007, **99**, 048102-048102.
14. J. Palacci, C. Cottin-Bizonne, C. Ybert and L. Bocquet, *Phys. Rev. Lett.*, 2010, **105**, 088304.
15. G. Volpe, S. Gigan and G. Volpe, *Am. J. Phys.*, 2014, **82**, 659-664.
16. E. M. Purcell, *Am. J. Phys.*, 1977, **45**, 3-11.
17. X. Chen, X. Yang, M. Yang and H. Zhang, *EPL*, 2015, **111**, 54002.
18. M. C. Marchetti, J.-F. Joanny, S. Ramaswamy, T. B. Liverpool, J. Prost, M. Rao and R. A. Simha, *Rev. Mod. Phys.*, 2013, **85**, 1143.
19. E. Lauga, *Annu. Rev. Fluid Mech.*, 2016, **48**, 105-130.
20. A. Furukawa, D. Marenduzzo and M. E. Cates, *Phys. Rev. E*, 2014, **90**, 022303.
21. E. Lauga and T. R. Powers, *Rep. Prog. Phys.*, 2009, **72**, 096601.
22. J. P. Hernandez-Ortiz, C. G. Stoltz and M. D. Graham, *Phys. Rev. Lett.*, 2005, **95**, 204501.
23. J. Elgeti, R. Winkler and G. Gompper, *Rep. Prog. Phys.*, 2015, **78**.
24. A. Baskaran and M. C. Marchetti, *Proc. Natl. Acad. Sci. U. S. A.*, 2009, **106**, 15567-15572.
25. T. Debnath, P. K. Ghosh, F. Nori, Y. Li, F. Marchesoni and B. Li, *Soft matter*, 2017, **13**, 2793-2799.
26. B.-Q. Ai and F.-G. Li, *Soft matter*, 2017, **13**, 2536-2542.
27. U. M. B. Marconi and C. Maggi, *Soft matter*, 2015, **11**, 8768-8781.
28. N. Koumakis, C. Maggi and R. Di Leonardo, *Soft matter*, 2014, **10**, 5695-5701.
29. A. Suma, G. Gonnella, D. Marenduzzo and E. Orlandini, *EPL*, 2014, **108**, 56004.
30. F. Peruani, A. Deutsch and M. Bär, *Phys. Rev. E*, 2006, **74**, 030904.
31. A. Kudrolli, G. Lumay, D. Volfson and L. S. Tsimring, *Phys. Rev. Lett.*, 2008, **100**, 058001.
32. S. R. McCandlish, A. Baskaran and M. F. Hagan, *Soft Matter*, 2012, **8**, 2527-2534.
33. W. F. Paxton, K. C. Kistler, C. C. Olmeda, A. Sen, S. K. St. Angelo, Y. Cao, T. E. Mallouk, P. E. Lammert and V. H. Crespi, *J. Am. Chem. Soc.*, 2004, **126**, 13424-13431.
34. M. Cates and J. Tailleur, *EPL*, 2013, **101**, 20010.
35. A. M. Menzel, *Phy. Rep.*, 2015, **554**, 1-45.
36. J. Stenhammar, A. Tiribocchi, R. J. Allen, D. Marenduzzo and M. E. Cates, *Phys. Rev. Lett.*, 2013, **111**, 145702.
37. G. Volpe and G. Volpe, *Am. J. Phys.*, 2013, **81**, 224-224.
38. T. T. Perkins, *Laser & Photon. Rev.*, 2009, **3**, 203-220.
39. G. Szamel, *Phys. Rev. E*, 2014, **90**, 012111.
40. J. Tailleur and M. Cates, *EPL*, 2009, **86**, 60002.
41. J. D. Weeks, D. Chandler and H. C. Andersen, *J. Chem. Phys.*, 1971, **54**, 5237-5247.
42. S. Plimpton, *J. Comput. Phys.*, 1995, **117**, 1-19.
43. T. Vicsek, A. Czirók, E. Ben-Jacob, I. Cohen and O. Shochet, *Phys. Rev. Lett.*, 1995, **75**,

1226.
44. K. Martens, L. Angelani, R. Di Leonardo and L. Bocquet, *Eur. Phys. J. E.*, 2012, **35**, 84.